\documentclass[aps, prb, reprint, lengthcheck, footinbib, showpacs, superscriptaddress]{revtex4-1}

\usepackage[T1]{fontenc}
\usepackage[english]{babel}

\usepackage{textcomp}
\usepackage{amsmath}
\usepackage{braket}
\usepackage{graphicx}

\usepackage[colorlinks, urlcolor=blue, citecolor=blue, linkcolor=blue, pdfstartview=FitH]{hyperref}
\usepackage[all]{hypcap}

\usepackage{times}
\usepackage{mathptmx}

\begin{document}

\title{Hanle effect in (In,Ga)As quantum dots: Role of nuclear spin fluctuations}

\author{M.~S.~Kuznetsova}
\affiliation{Spin Optics Laboratory, St.~Petersburg State University, 198504 St.~Petersburg, Russia}
\author{K.~Flisinski}
\affiliation{Experimentelle Physik 2, Technische Universit\"at Dortmund, D-44221 Dortmund, Germany}
\author{I.~Ya.~Gerlovin}
\author{I.~V.~Ignatiev}
\affiliation{Spin Optics Laboratory, St.~Petersburg State University, 198504 St.~Petersburg, Russia}
\author{K.~V.~Kavokin}
\affiliation{Spin Optics Laboratory, St.~Petersburg State University, 198504 St.~Petersburg, Russia}
\affiliation{A.~F. Ioffe Physical-Technical Institute, Russian Academy of Sciences, 194021 St.~Petersburg, Russia}
\author{S.~Yu.~Verbin}
\affiliation{Spin Optics Laboratory, St.~Petersburg State University, 198504 St.~Petersburg, Russia}
\author{D.~R.~Yakovlev}
\affiliation{Experimentelle Physik 2, Technische Universit\"at Dortmund, D-44221 Dortmund, Germany}
\affiliation{A.~F. Ioffe Physical-Technical Institute, Russian Academy of Sciences, 194021 St.~Petersburg, Russia}
\author{D.~Reuter}
\author{A.~D.~Wieck}
\affiliation{Angewandte Festk\"orperphysik, Ruhr-Universit\"at Bochum, D-44780 Bochum, Germany}
\author{M.~Bayer}
\affiliation{Experimentelle Physik 2, Technische Universit\"at Dortmund, D-44221 Dortmund, Germany}

\date{\today}

\begin{abstract}
The role of nuclear spin fluctuations in the dynamic polarization of
nuclear spins by electrons is investigated in (In,Ga)As quantum dots. The
photoluminescence polarization under circularly polarized optical
pumping in transverse magnetic fields (Hanle effect) is studied. A weak
additional magnetic field parallel to the optical axis is used to
control the efficiency of nuclear spin cooling and the sign of nuclear
spin temperature. The shape of the Hanle curve is drastically modified
with changing this control field, as observed earlier in bulk
semiconductors and quantum wells. However, the standard nuclear spin
cooling theory, operating with the mean nuclear magnetic field
(Overhauser field), fails to describe the experimental Hanle curves in
a certain range of control fields. This controversy is resolved by taking into
account the nuclear spin fluctuations owed to the finite number of
nuclei in the quantum dot. We propose a model describing cooling of
the nuclear spin system by electron spins experiencing fast vector
precession in the random Overhauser fields of nuclear spin fluctuations.
The model allows us to accurately describe the measured Hanle curves
and to determine the parameters of the electron-nuclear spin system of
the studied quantum dots.
\end{abstract}

\pacs{78.67.Hc, 78.47.jd, 76.70.Hb, 73.21.La}

\maketitle

\section*{Introduction}

The hyperfine interaction of electron spins with the spins of lattice nuclei is able
to create a considerable dynamic nuclear polarization (DNP) in semiconductors
under optical pumping by circularly polarized light.\cite{OO} In this process,
the angular momentum received by the electron from a photon is transferred
to the nuclear spin system. In turn, the magnetic moment of polarized nuclei
affects the electron spin as an effective magnetic field (Overhauser field),
giving rise to splitting of electron spin states. Under conditions of strong optical
pumping the splitting can reach tens of micro-eV so that the nuclear polarization
becomes detectable, in certain cases, by high-resolution optical spectroscopy.
\cite{BrownPRB96, GammonPRL01, KrebsPRL10, CherbuninPRB09}

An alternative approach to detection of nuclear polarization, which does not
require high spectral resolution, is to measure the electron polarization created
by optical pumping in an external magnetic field. As the non-equilibrium electron spin polarization is in many cases magnetic-field dependent, the Overhauser field can be detected using its effect on the mean electron spin, for example by observing the associated changes in the circular polarization of photoluminescence (PL).\cite{OO}
In a magnetic field parallel to the optical axis (longitudinal magnetic field), the nuclear polarization created by the
pumping may influence the PL polarization by suppressing electron spin relaxation \cite{OO, CherbuninPRB09}. For optical pumping in a magnetic field perpendicular to the optical axis, the electron spin polarization is usually destroyed with increasing magnetic field (Hanle effect). In this case the Overhauser field modifies the width and shape of the dependence of the circular polarization of the PL on the magnetic field (Hanle curve), which can become non-monotonous and even hysteretic \cite{OO, CherbuninPRB09, KalevichSpinBook,
PagetPRB77, PalPRB09, AuerPRB09, FlisinskiPRB10,CherbuninPRB11, MasumotoPRB08, UrbaszekRMP12}

The Hanle effect in presence of nuclear spin polarization in bulk semiconductors and quantum wells has been theoretically treated in the model of mean Overhauser field, which has provided good qualitative and quantitative agreement with experimental data. The validity of the mean-field approach in these systems is justified by the fact that the correlation time $\tau_c$ of the electron spin at the position of a certain nucleus is much shorter than the electron's spin lifetime $T_s$. Indeed, electrons localized at shallow impurity centers or structural imperfections rapidly loose their spin polarization to other, localized or itinerant, electrons via exchange scattering \cite{PagetExchange, KavokinSST08}; as this process is spin-conserving, the mean polarization of the entire electron ensemble lives over a much longer time scale determined by spin-orbit or hyperfine interactions. As a result, the fluctuations of the Overhauser field $B_N$ are effectively averaged out and give rise only to the electron spin relaxation, which is in this case exponential, with the decrement $\tau_s^{-1}={\gamma_e}^2(\langle{B_N}^2\rangle-{\langle{B_N}\rangle}^2)\tau_c $, where $\gamma_e$ is the electron gyromagnetic ratio \cite{OO}. This approach, called approximation of short correlation time \cite{KalevichSpinBook}, often fails in quantum dots where electron states are strongly localized and effectively isolated from all the other electrons. In this case, the electron spin is exposed to a virtually static nuclear spin fluctuation (NSF) during its entire lifetime (note that the correlation time of nuclear spins, which is of the order of their transverse relaxation time $T_2\approx10^{-4}s$, is orders of magnitude longer than typical electron spin lifetimes).\cite{MerkulovPRB02, KhaetskiiPRL02,  MerkulovPRB10} The Larmor precession of the electron spin in the fluctuation nuclear field was predicted to result in a specific non-exponential pattern of electron spin decay \cite{MerkulovPRB02}, which was subsequently experimentally observed.\cite{BraunPRL05} The influence of NSF on the evolution of the regular Overhauser field and, eventually, on the Hanle effect under dynamic polarization of nuclear spins has so far not been studied experimentally.

In this paper, we report on detailed measurements of the Hanle effect in (In,Ga)As/GaAs QDs in the weak-field range (0-20 mT field strength), where the effect of the NSF is expected to be the strongest. We have measured a set of Hanle curves under optical excitation of moderate intensity and at different strengths of an additional magnetic field applied along the optical axis (longitudinal magnetic field). The experimental Hanle curves are compared with the results of calculations using two models, one including NSF and the other one taking into account only mean Overhauser fields. In both theories, the mean Overhauser field has been calculated within the spin temperature approach.\cite{OO} Our analysis shows that the mean-field model fails to describe the features of the Hanle curve around zero transversal field where the so-called W-structure appears in a certain range of longitudinal fields $B_z$. The model including NSF, on the other hand, yields good fits of the experimental data, with a reasonable choice of parameters, for all experimental conditions but for the exact compensation of the Knight field with $B_z$. In the latter case, nuclear quadrupole effects due to strain in the QDs probably play the dominant role.

\section*{Experimental details}

A heterostructure containing 20 layers of self-assembled (In,Ga)As/GaAs QDs
separated by Si-$\delta$-doped GaAs barriers was studied. The heterostructure was annealed at temperature 980~\textdegree{}C
which resulted in the considerable decrease of mechanical stress in the QDs and in enlarging
the localization volume for resident electrons due to inter-diffusion of Ga and In atoms.
The sample was immersed in liquid helium at $T=1.8~K$ in a cryostat with a superconducting magnet.
Magnetic fields up to $100$~mT were applied perpendicular to the optical axis (Voigt geometry) along to
the $\lbrack110\rbrack$ crystallographic direction of the sample. To create an additional magnetic field, perpendicular
to the main magnetic field and parallel to the optical axis, a pair of small Helmholtz coils was installed outside the cryostat.

The PL of the sample is excited by circularly polarized light from a continuous-wave Ti:sapphire laser,
with the photon energy tuned to the optical transitions in the wetting layer of the sample.
The degree of circular polarization of the PL is detected by a standard method using a photoelastic
modulator and an analyzer (a Glan-Thompson prism). The modulator creates a phase difference, $\Delta\varphi=(\pi/4)\sin(2\pi ft)$,
between the linear components of the PL thus converting each of the circular components ($\sigma^+$ and $\sigma^-$)
into linear ones ($x$ and $y$) at frequency $f = 50$~ kHz. The analyzer selects one of the linear
components, which was dispersed with a 0.5-m monochromator and detected by an avalanche photodiode.
The signal from the photodiode was accumulated for each circular component separately in a two-channel
photon counting system. The PL polarization was recorded at the wavelength corresponding to the maximum
of PL band of the sample. Typical polarization-resolved PL spectra for the sample under study can be found in Ref.~\onlinecite{FlisinskiPRB10, VerbinJETF12}.

The degree of PL polarization of the QDs is negative, i.e., the PL is predominately
$\sigma^-$-polarized for $\sigma^+$-polarized excitation. The mechanism of negative circular polarization (NCP)
has been extensively discussed in Ref.~\onlinecite{CortezPRL02, ShabaevPRB09, IgnatievOS09} where it was shown that the presence of NCP is the result
of optical orientation of the resident electrons provided by ionization of donors outside the QDs (the doping level of our structure corresponds to one resident electron per QD on average). The amplitude of NCP is proportional to the projection of electron spin onto the optical axis $z$, averaged over the QD ensemble\cite{CherbuninPRB11}:
\begin{equation}
A_{NCP}\sim2\braket{S_z}.
\label{eq:Amplitude}
\end{equation}

The amplitude of the central peak increases with rising excitation power at relatively low excitation levels. 
A further rise of the power results in saturation of the peak amplitude which indicates a high level of electron spin polarization. 
The pump powers used in our experiments were sufficient to totally polarize the electron spin.]

In this paper we use the absolute value of NCP as a measure of the electron spin orientation.
We studied the dependence of PL polarization on magnetic field applied perpendicular
to the optical axis. The central part of the Hanle curve, where the W-like structure is
observed, was studied most carefully, to understand the role of the Knight field in
the optical orientation of nuclear spins. In particular, modifications of the W-structure
under application of small magnetic fields parallel to the optical axis was studied.

\section*{Experimental results}

\begin{figure}[h]
\includegraphics[clip,width=.85\columnwidth]{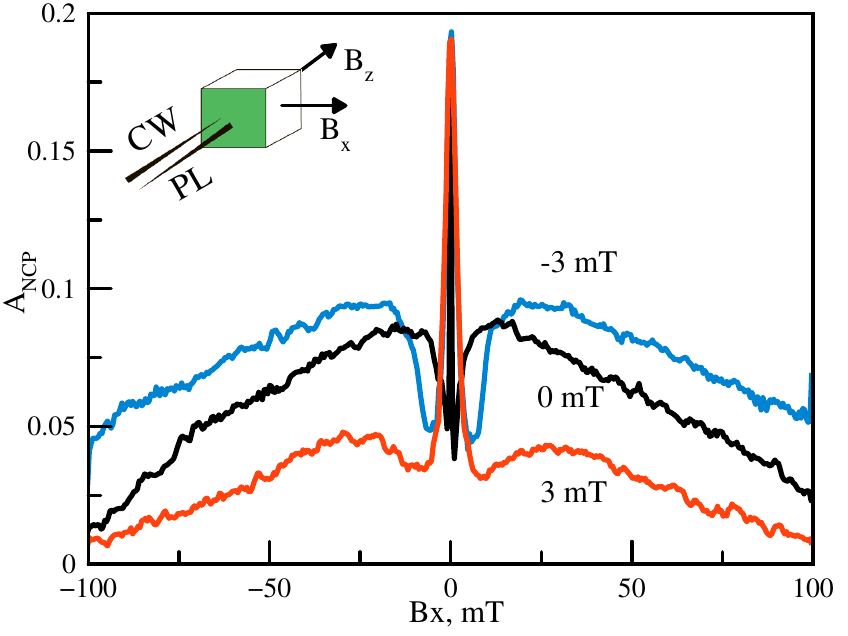}
\caption{Overall shape of Hanle curves measured at
different longitudinal magnetic fields $B_z$ indicated at each curve. Inset shows configuration of the experiment.}
\label{fig:One}
\end{figure}

The general form of Hanle curves measured in absence of longitudinal magnetic fields,
$B_z$, as well as in presence of small $B_z$ is shown in Fig.~\ref{fig:One}. Each curve shows the
pronounced W-structure consisting of the narrow central peak and two maxima
positioned symmetrically relative to the peak.  
 Application of $B_z$ considerably affected the central peak width as well as the overall width of the Hanle curve.

Namely, the central-peak width
increases with $B_z$ irrespectively of its sign. At the same time, the width of the Hanle curve
drops approximately twice when $B_z$ is changed from $-3$~mT to $+3$~mT. This difference
in behavior of the Hanle curve and its central peak is an indication that they are controlled
by different components of the hyperfine interaction. According to Ref.~\onlinecite{OO}, the width of
the central peak and the shape of the W-structure are determined by the dynamic nuclear polarization
creating an effective field parallel to the external magnetic field. The large width of the Hanle
curve is due to partial stabilization of the electron spin orientation along the optical
axis by the longitudinal component of the effective nuclear field of quadrupole-split nuclear spin states.\cite{DzhioevPRL07} 
The analysis of this effect and the variation of the Hanle curve width with $B_z$ requires a separate study; we consider hereafter
only the behavior of the central part of the Hanle curve.

The effect of longitudinal magnetic fields, ranging from
 $-20$~mT to $+20$~mT, on the Hanle curve is shown in Fig.~\ref{fig:Two}. The experiment shows that
application of a positive $B_z$ is accompanied by a monotonous increase of
the width of the central peak and of the dips near the peak. The depth of the
dips remains almost unchanged. At negative  $B_z$, the behavior of the dips is
not monotonous. The change of $B_z$ from zero to $-1$~mT results in
almost total disappearance of the dips without noticeable change of their width.
A further increase of $B_z$ in absolute value leads to the increase of both depth
and width of the dips. At the same time, the width of the central peak monotonously
rises with increasing absolute value of  $B_z$.

\section*{Analysis}

\section*{Standard nuclear spin--cooling model}

First, we analyze the experimental data presented above, using
the standard mean-field model of nuclear spin cooling, applied for the first time to the analysis
of the w-structure of Hanle curves in Ref.~\onlinecite{PagetPRB77}.
The presence of well-resolved W-structures in the Hanle curves is a clear
indication of nuclear polarization created by optical pumping.
The nuclear spin-cooling model \cite{DPcooling, PagetPRB77, OO} assumes that the electron spin orientation maintained by optical excitation is the
source of a permanent flux of angular momentum into the nuclear spin system, which creates a dynamical nuclear polarization.
Due to the fast transverse relaxation, only the component of nuclear spin momentum parallel to the total effective magnetic field acting on the nuclei is conserved. The effective field is the sum of the external magnetic field and of the Knight field created by hyperfine interaction with the polarized electron spin. Nuclear spin orientation along the magnetic field changes the Zeeman energy of nuclear spins, which results in a change of their spin temperature. In the case of permanent optical pumping, this change is determined by the balance between the energy flux into the spin system and the energy flux from the nuclei into the phonon bath due to longitudinal spin relaxation. The spin temperature controls the magnitude of accumulated nuclear spin momentum\cite{Abragam} and, therefore, the effective field acting on the electron spin. Thus the dynamics of electron spin is determined by the joint action of the external and nuclear fields, while the dynamics of the nuclear polarization depends, in turn, on the Knight field created by the polarized electron spin. The nuclear polarization affects the electron spin via the Overhauser field, which is parallel or anti-parallel to the nuclear spin depending on the sign of the electron g-factor. In particular, it is anti-parallel for positive sign of $g_e$.\cite{OO}
It is essential that the efficiency of nuclear spin cooling is proportional to the scalar product of the electron spin times the total field acting on the nuclei. When the external magnetic field is strictly perpendicular to the optical axis, the nuclear spin cooling occurs only due to the Knight field.

The above consideration allows one to derive a system of coupled balance equations
for the electron spin and the nuclear spin temperature. Its steady state solution yields a cubic equation for the average projection of electron spin onto the direction of
observation. This equation for a magnetic field, $B_x$, perpendicular to the optical
axis, is given in Ref.\onlinecite{OO}. Simple generalization of the equation is possible for the case when
an additional magnetic field, $B_z$, directed along the optical axis is present:
\begin{equation}
S_z\left(1+\frac{K^2}{B^2_{1/2}}B^2_x\right)-S_0\left(1+\frac{K^2}{B^2_{1/2}}B^2_z\right)=0,
\label{eq:Eq1}
\end{equation}
where
\begin{equation*}
K=1+\frac{S_0 B_z+b_e S_0 S_z}{B^2_x+2 b_e S_0 B_z+b^2_e S_0 S_z+\xi B^2_L}.
\end{equation*}
Here $S_0$ is the initial electron spin orientation created by excitation, $S_z$ is the $z$--projection of the electron spin averaged
over time , $B_{1/2}=\pi/{(\mu_B g_e T_e)}$  where $g_e$, $\mu_B$ and $T_e$ is the electron $g$--factor, the Bohr
magneton and the electron spin life time, respectively, $b_e$ is the coefficient of proportionality between the electron spin and the Knight field.
Term $\xi B^2_L$ in Eq. ~\eqref{eq:Eq1}  describes the interaction between nuclear spins causing the relaxation
of nuclear polarization, where $B_L$ is the local field of the nuclear dipole-dipole interactions.

\begin{figure*}
\includegraphics[clip,width=.85\columnwidth]{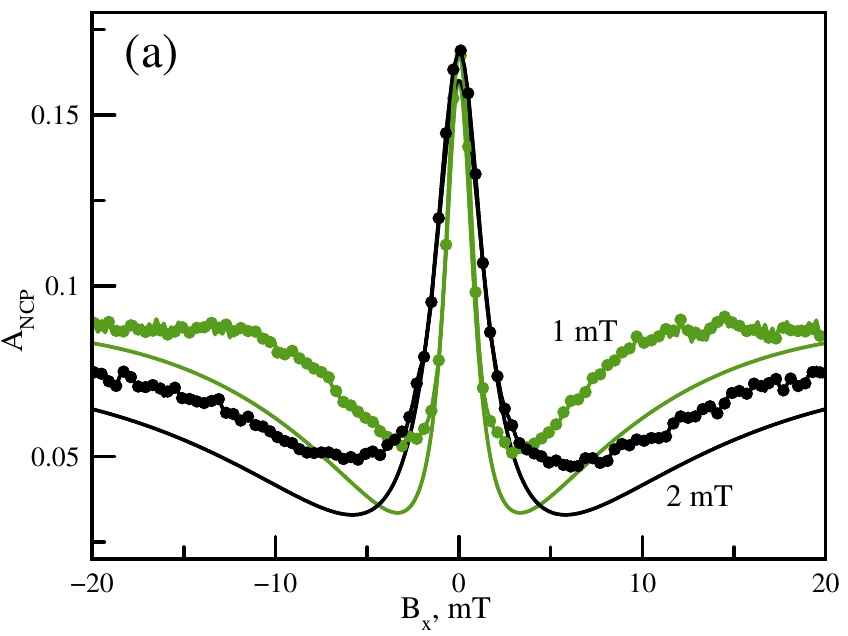}
\includegraphics[clip,width=.85\columnwidth]{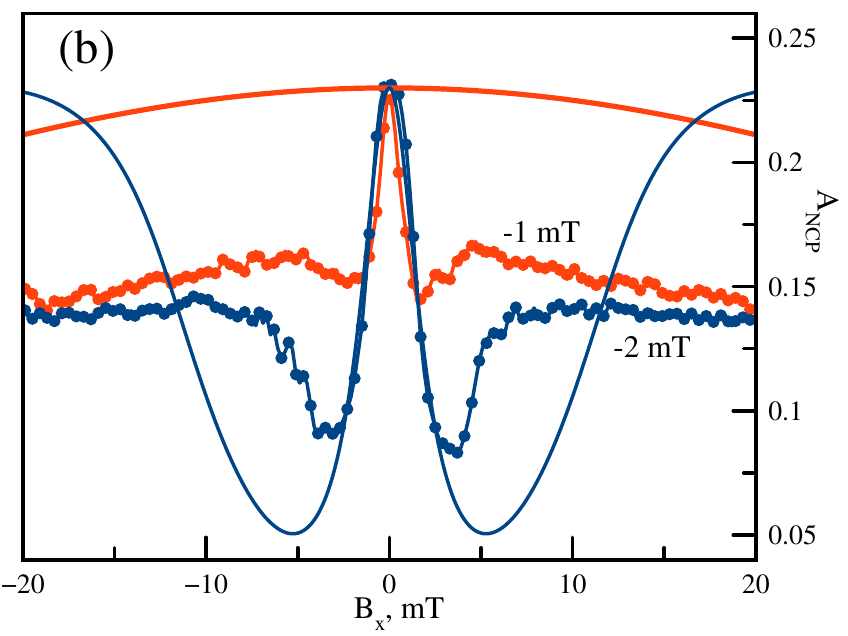}
\caption{Results of calculations in the framework of the standard cooling model (solid lines) in comparison with the experimental data (points) for
(a) negative and (b) positive longitudinal external fields, $B_z$. Values of $B_z$ are given at each curve.}
 \label{fig:Two}
\end{figure*}

For modeling of the experimentally measured Hanle curve, we have numerically
solved equation~\eqref{eq:Eq1} and obtained $S_z$ as a function of the transverse magnetic
field $B_x$ for different values of the longitudinal magnetic field $B_z$ in the range from $-3$~mT to $+3$~mT.
The following values of the other parameters were used in the calculations: $S_0=1/2$, $B_{1/2}=60$~mT, $B_L=0.3$~mT, $b_e=2.0$~mT.
Most of them approximately correspond to our experimental conditions and the properties of the sample studied.
The exception is the value of $B_{1/2}$ extracted from the experimentally measured width of the Hanle curve.
It corresponds to an electron spin life time of the order of $10^{-10}~s$, which is several orders of magnitude smaller
than the real value in the structures of this type (see, for example, Ref.\onlinecite{OultonPRL07}).

Examples of the calculated dependences are shown in  Fig.~\ref{fig:Two}. They indeed demonstrate behaviors similar to the measured Hanle curves. This is in particular 
true for positive $B_z$ (see Fig.~\ref{fig:Two}~(b)), which, for the helicity of excitation used in our experiments, is co-directed to the Knight field. The analysis
shows that the effective nuclear field in this case is co-directed to the external magnetic field
 and thus ``amplifies'' it. This amplification results in gradual decrease of
 spin polarization and, correspondingly, of PL polarization with rising $B_z$ beyond the central peak, as seen
 both from the calculations and from the measured curves.

When $B_z$ is negative, the effective field is anti-parallel to the Knight field, $B_e = S_0b_e$. If $B_z = -B_e$, so that
the compensation of the longitudinal component of total field occurs. According to Refs.~\onlinecite{OO, PagetPRB77},
nuclear spin cooling is not possible in this case. This should result in disappearance of the
W-structure, as it is indeed seen in Fig.~\ref{fig:Two}~(a) for the Hanle curve calculated for $B_z = -1$~mT.
 At more negative $B_z$, the W-structure appears again, but the additional maxima run away form the central
peak with increasing $| B_z |$, maintaining the same amplitude as the central peak. This behavior of the calculated Hanle
 curves is explained by the fact that in this case the nuclear field is directed against the total effective magnetic field
affecting the nuclei. The $x$-component of the nuclear field, $B_{Nx}$, is compensated by the transverse magnetic field $B_x$
at some magnitude of $B_x$, giving rise to the additional maxima. The efficiency of the nuclear-spin pumping
increases with increase of $|B_z|$. As a result, $B_{Nx}$ increases, and the positions of the compensation points where
$B_{Nx} + B_x = 0$ are shifted to larger $|B_x|$.

These numerical results, however, are in strong contradiction with our
experimental observations [see Fig.~\ref{fig:Two}~(a)]. The central peak of the measured Hanle curves is higher than the other
parts of the Hanle curve at any negative $B_z$. We want to stress that the disagreement between
the theory and the experiment cannot be eliminated for any set of values of the adjustable parameters.
Therefore this contradiction is of principal importance and indicates that the model of mean nuclear field ignores
some mechanism causing depolarization of the electron spin at non-zero transverse magnetic field, including points where it is totally compensated by the nuclear field.
The discrepancy is evident also from the unrealistically large value of $B_{1/2}$ needed to fit, at least partly, the experimental Hanle curves within the mean-field model.

\section*{Effect of nuclear spin fluctuations}

 To extend the standard cooling model in order to account for the effects of NSF, we suppose that the effective
nuclear field consists of a regular component, $\mathbf{B}_{N}$, created by the nuclear polarization, and a fluctuating
component, $\mathbf{B}_{f}$, appearing due to the random orientation of the limited number of nuclear spins interacting
 with the electron spin.\cite{MerkulovPRB10} The estimates given in Refs.~\onlinecite{PalPRB09, PetrovPRB08} for similar QDs show that the average magnitude
 of the fluctuating nuclear field is of the order of tens of milliTesla. The frequency of electron spin precession about the
 field is orders of magnitude larger than the rate of longitudinal relaxation of the electron spin. Therefore, only the
 projection of electron spin onto the field is conserved. The magnitude and the direction of the fluctuating field are randomly
 distributed in the QD ensemble. In the absence of other fields, such as the external magnetic field and the field
of nuclear polarization, the depolarization of the electron spin by the fluctuating field reduces the observable $z$--component
 of spin polarization to $1/3$ of its initial value.\cite{MerkulovPRB02, PetrovPRB08}

An effective optical pumping can create a dynamic nuclear polarization, whose magnitude
can considerably exceed the nuclear spin fluctuations. If the transverse magnetic field is zero,
the effective field of nuclear polarization is directed along the optical axis and is able to suppress the
effect of NSF. This results in the increased amplitude of the central peak of the Hanle curve. Experiments
show~\cite{CherbuninPRB11, IkezawaPRB05} that the PL polarization at zero transverse magnetic field can reach 50\% or even more.
When $B_x \neq 0$, the nuclear polarization deviates from the optical axis, and its $z$-component decreases, so
that the NSF can reduce the electron spin polarization. In particular, the electron spin
polarization at the point of mutual compensation of the external field and the field of nuclear polarization
is smaller than the polarization at zero $B_x$. This qualitative consideration explains the small amplitudes of the
additional maxima of the Hanle curves, which cannot be explained by the mean-field model.

In order to include NSF in the theory, we use the fact that the build-up time of the nuclear polarization is much longer than the correlation time of the nuclear spin fluctuation ($\approx{T_2}$), which is, in turn, orders of magnitude longer than the electron spin lifetime. For this reason, the nuclear spin temperature can be calculated using the value of the electron mean spin averaged over possible realizations of the NSF, while each NSF realization can be considered "frozen" (i.e. the evolution of nuclear spin during the electron spin lifetime can be neglected~\cite{MerkulovPRB02}). The dependence of the average electron spin polarization on the transverse external magnetic field within this approximation
is a bell-like curve, which can be well fitted by a Lorentzian:
\begin{equation}
\rho (B_x)\approx\frac{\braket{B_{fz}^2}}{B^2+\braket{B^2_f}},
\label{eq:Eq3}
\end{equation}
Here $\braket{B_f^2}=\braket{B^2_{fx}}+\braket{B^2_{fy}}+\braket{B^2_{fz}}$,
where
$\braket{B^2_{f \alpha}}$
is the squared $\alpha$-component ($\alpha = x, y, z$) of the NSF averaged over the QD ensemble. Eq.~\eqref{eq:Eq3}
has a simple geometrical interpretation. In each QD with realization of a particular fluctuating field, $\mathbf{B}_f$,
only the projection of the electron spin onto the total field, $\mathbf{B}^{(e)}_{tot}=\mathbf{B}_x+\mathbf{B}_f$,
survives: $S_\parallel =S_0\cos\phi$, where $\varphi$ is the angle between the vector $\mathbf{B}^{(e)}_{tot}$  and the
$z$-direction. The experimentally observable quantity $\rho (B_x)$  is proportional to the $z$-projection of the electron spin, $S_z=S_\parallel\cos\varphi=S_0\cos^2\varphi$, where $S_0 = 1/2$. It is obvious that in such conditions $\cos^2\varphi={B^2_{fz}}/\left(B^{(e)}_{tot}\right)^2$. Averaging over the QD ensemble gives rise to Eq.~\eqref{eq:Eq3} if we neglect the correlations of the quantities in the numerator and the denominator of this equation.

Some generalization of Eq.~\eqref{eq:Eq3} is required to describe electron spin polarization under our experimental conditions.
We need to take into account the regular nuclear field, $\mathbf{B}_{N}$, with non-zero components $B_{Nx}$ and $B_{Nz}$ created
by the dynamic polarization of nuclei. Similar to the standard mean-field model, we assume for simplicity that the
the electron density is homogeneously distributed over the nuclei
(the so-called ,,box'' model  approximation\cite{KozlovJETF07}), which allows us to neglect the spatial variation of the Knight field. Also, since we consider weak magnetic fields, we describe all the nuclear species with a single spin temperature~\cite{OO}. Since the effective field of nuclear polarization has a certain direction
 (in contrast to the NSF field) its components are either added to or subtracted from the respective components of
the external magnetic field, depending on the experimental conditions. In addition, in our experiments, the external
 magnetic field has not only the transverse component but also some longitudinal one. For this case we can write down
the following expressions for the $z$- and $x$-components of the averaged electron spin, $S_\parallel$:
\begin{equation}\label{eq:Eq4}
S_z=S_0\frac{(B_z+B_{Nz})^2+\braket{B^2_{fz}}}{(B_x+B_{Nx})^2+(B_z+B_{Nz})^2+\braket{B^2_f}}.
\end{equation}
\begin{equation}\label{eq:4}
S_x=S_0\frac{(B_z+B_{Nz})(B_x+B_{Nx})}{(B_x+B_{Nx})^2+(B_z+B_{Nz})^2+\braket{B^2_f}} \tag{\ref{eq:Eq4}$'$}
\end{equation}
Here we assume that the regular nuclear field $\mathbf{B}_N$ is directed along the total effective field,$\mathbf{B}^{(N)}_{tot}$, acting on the nuclei,
 which consists of the external magnetic field, $\mathbf{B}_x +\mathbf{B}_z$, and the Knight field, $\mathbf{B}_e = b_e\mathbf{S_\parallel}$, created by hyperfine
 interaction with the electron spin. According to the standard cooling model\cite{OO}, the nuclear field $\mathbf{B_N}$ is determined by the equation:
\begin{equation} \label{eq:Eq5}
\mathbf{B_N}=\mathbf{B^{(N)}_{tot}}\frac{b_N(\mathbf{B^{(N)}_{tot}}\bullet\mathbf{S_\parallel})}{B^{(N)^2}_{tot}+\xi B^2_L}\bullet 4I(I+1)/3
\end{equation}
where the parameter $b_N$ is the effective field of totally polarized nuclei affecting the electron spin.
The magnitude of $b_N$ is determined by the properties of the particular electron-nuclear spin system
 and should not depend on external conditions. The factor $4I(I+1)/3$ can be omitted because
 we consider only the nuclear field created by spin states with $I = \pm1/2$.

The above equation allows one to obtain the following expressions for the $x$-- and $z$--components of the nuclear field:
\begin{equation}\label{eq:5}
B_{Nx}=(B_x+b_e S_x)\frac{b_N(B_z S_z+B_x S_x+b_e S^2_x+b_e S^2_z)}{(B_x+b_e S_x)^2+(B_z+b_e S_z)^2+\xi B^2_L}, \tag{\ref{eq:Eq5}$'$}
\end{equation}
\begin{equation}\label{eq:5!}
B_{Nz}=(B_z+b_e S_z)\frac{b_N(B_z S_z+B_x S_x+b_e S^2_x+b_e S^2_z)}{(B_x+b_e S_x)^2+(B_z+b_e S_z)^2+\xi B^2_L} \tag{\ref{eq:Eq5}$''$}
\end{equation}
The coefficient $b_e$ is given, in principle, by\cite{OO}: $b_e=-(16\pi/3)\mu_B\zeta^2$ , where $\mu_B$ is the Bohr magneton
and $\zeta$ is the electron density on a nuclear site. Negative sign means that the direction of the Knight
field is \emph{opposite} to that of the electron spin. Because the electron density is dependent on the QD size,
which can sufficiently vary from dot to dot, the value of $\zeta$ is unknown \emph {a priory}.

Equations \eqref{eq:Eq4}, \eqref{eq:4}, \eqref{eq:5}, and \eqref{eq:5!} contain the Cartesian components of the electron spin
 and of the dynamic nuclear polarization as unknown quantities. We found them by numerical solution of these equations for
 transverse magnetic fields in the range $-20 \cdots +20$~mT and  for the several values of the longitudinal magnetic field used in experiment.

In the calculations, the coefficient $b_e$ has been chosen such that the Knight field compensates the $z$-component of the magnetic
 field at the point where the dips near the central peak of the Hanle curve disappear (see Fig.~\ref{fig:Three}). The quantities $b_N$ , $B_{fz}$ and $\xi B^2_L$ were
 considered as fitting parameters and varied to get the best correspondence with the experimentally obtained Hanle curves.
 To compare the calculated results with experimental data we multiplied the calculated values of $S_z$ by a factor $\alpha$, which takes into
 account the reduced magnitude of PL polarization. This reduction is presumably due to the fact that some QDs are charge neutral and their PL
is non-polarized.  $\alpha = 0.2\pm 0.02$ for curves measured at negative $B_z$ and $0.16\pm 0.01$ for positive values of $B_z$.
 The latter curves were measured at slightly lower power of excitation. 
The possible reason for the pump-power dependence of ${\alpha}$ is the creation of photo induced electrons which slightly change the fraction of charged QD's.

We should note that equations~\eqref{eq:Eq4}, \eqref{eq:4}, \eqref{eq:5}, and \eqref{eq:5!} are inter-connected cubic equations. Their solution is unstable in the most general case, complicating the determination of fitting parameters. To simplify the calculations, we performed them into two steps. In the first one, we excluded the $x$-component of electron spin from the equations because it weakly affects the nuclear polarization. Besides, we neglected the small difference in orientation of effective fields $\mathbf {B^{(e)}_{tot}}$ and $\mathbf {B^{(N)}_{tot}}$, and also introduced a fitting parameter:
\begin{equation}
b^{'}_N=b_N (B^N_{tot})^2/ \lbrack (B^N_{tot})^2+\xi B^2_L\rbrack,
\label{eq:Eq6}
\end{equation}
which characterizes the real nuclear field acting on the electron spin. This reduces the system of equations to one equation of fifth order for $S_z$:
\begin{figure*}
\includegraphics[clip,width=.85\columnwidth]{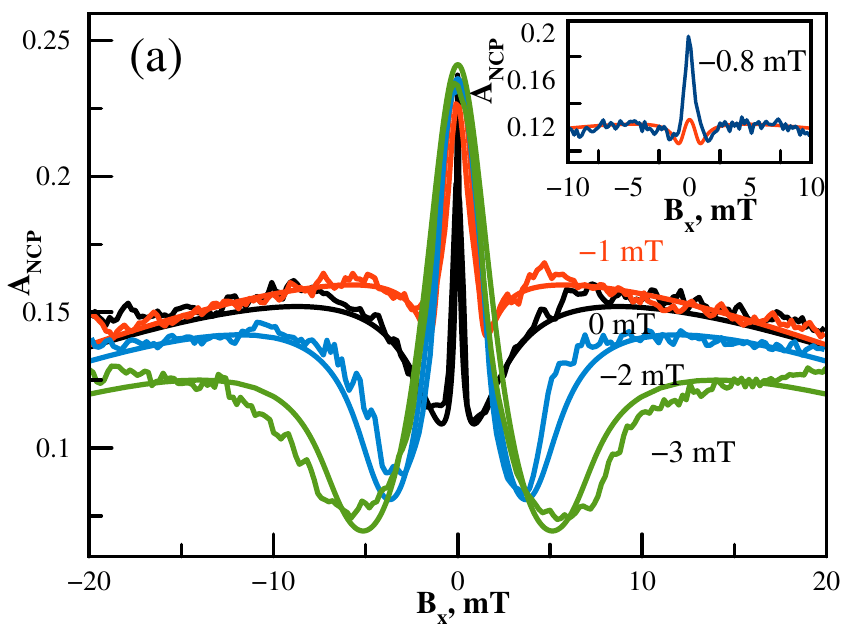}
\includegraphics[clip,width=.85\columnwidth]{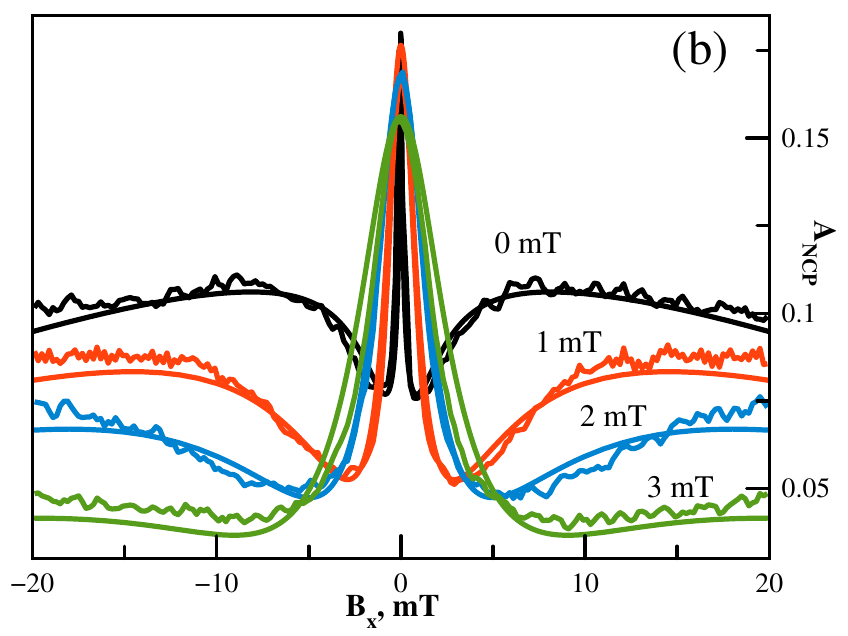}
\caption{Results of calculations taking into account the NSF (smooth solid lines) in comparison with experimental data (noisy lines) for negative (panel a) and positive (panel b)  longitudinal external fields, $B_z$. Values of $B_z$ are given near each curve. Inset shows central parts of calculated and experimental Hanle curves in the case of mutual compensation of $B_z$ and $B_e$.}
 \label{fig:Three}
\end{figure*}
\begin{widetext}
\begin{equation}
\rho =\frac{S_z}{S_0}=\frac{2B_{ez}}{b_e}=\frac{\left[B_zB_x^2+(B_z-S_0b^{'}_N)(B_{ez}+B_z)^2\right]^2+\braket{B_{fz}^2}\left[B_x^2+(B_{ez}+B_z)^2\right]}{B_x^2\left[B_x^2+(B_{ez}+B_z)^2-S_0b^{'}_N(B_{ez}+B_z)\right]^2+\left[B_zB_x^2+(B_z-S_0b^{'}_N)
(B_{ez}+B_z)^2\right]^2+\braket{B_{f}^2}\left[B_x^2+(B_{ez}+B_z)^2\right]}
\label{eq:Eq7}
\end{equation}
 \end{widetext}
We solved equation~\eqref{eq:Eq7} numerically, which allowed us to determine the range of possible values for quantities
 $b_N$ and $B_{fz}$. In the second step, we solved the whole system of equations~\eqref{eq:Eq4}, \eqref{eq:4}, \eqref{eq:5}, and \eqref{eq:5!} and used their real
roots for modeling the Hanle curves, slightly varying the fitting parameters determined in the first step.

Examples of the calculated Hanle curves are shown in Figs.~\ref{fig:Three}.  As seen there, reasonable agreement between calculated and measured curves is observed for positive as well as for negative $B_z$. Some deviations from the experiment occur for magnetic fields $B_z$ in the range $-0.5 .. -1$~mT, where the theoretically calculated amplitude of the central peak is considerably smaller than the one observed experimentally (see inset in Fig.~\ref{fig:Three}). The strong decrease of the peak amplitude obtained in the calculations is due to the depolarization of the electron spin by the nuclear spin fluctuations when the longitudinal component of total field  disappears and the nuclear field does not build up. This decrease is not observed experimentally. We have to assume that there is an additional effect of stabilization of electron spin in this case. A possible reason for that could be polarization of quadrupole-split nuclear spin states. Further studies of the quadrupole effects are needed to clarify this point.

The results of the calculations allow us to conclude that the effect of nuclear spin fluctuations is indeed important for the QDs under study. The good agreement between theory and experiment for the whole range of $B_z$ (with the only exception mentioned above) allows us to consider in more detail the physics meaning of the parameters obtained from the fitting and their dependence on the longitudinal magnetic field.

We find that the NSF amplitude, $\sqrt{\braket{B_{f}^2}}$, can be chosen close to $25$~mT for all the Hanle curves measured at various longitudinal magnetic fields.
This value is somewhat larger than the one obtained in another experiment with similar QDs.\cite {CherbuninPRB11} A possible reason for this overestimation of the NSF
amplitude is the increase ofthe  wings of the Hanle curves due to polarization of quadrupole--split nuclear spin states, which becomes noticeable at
 magnetic fields  $|B_x|\sim 20$~mT and larger.\cite{VerbinJETF12} By adding this polarization phenomenologically into the model, we were able to reduce the calculated
 NSF amplitude. We ignore here the quadrupole effects to avoid a complication of the analysis.

We have also verified the validity of the assumption of an isotropic distribution
of NSF replacing $\braket{B_{fz}^2}\rightarrow\beta\braket{B_{fz}^2}$   in the numerator of Eq.~\eqref{eq:Eq7} and optimizing the factor $\beta$.
The optimal value of $\beta$ was found to be in the range from $1.2$ to $1.4$. We suppose that some asymmetry of the distribution of nuclear spin fluctuations can also be due to the quadrupole stabilization of nuclear spins along the growth axis.

\begin{figure}[h]
\includegraphics[clip,width=.85\columnwidth]{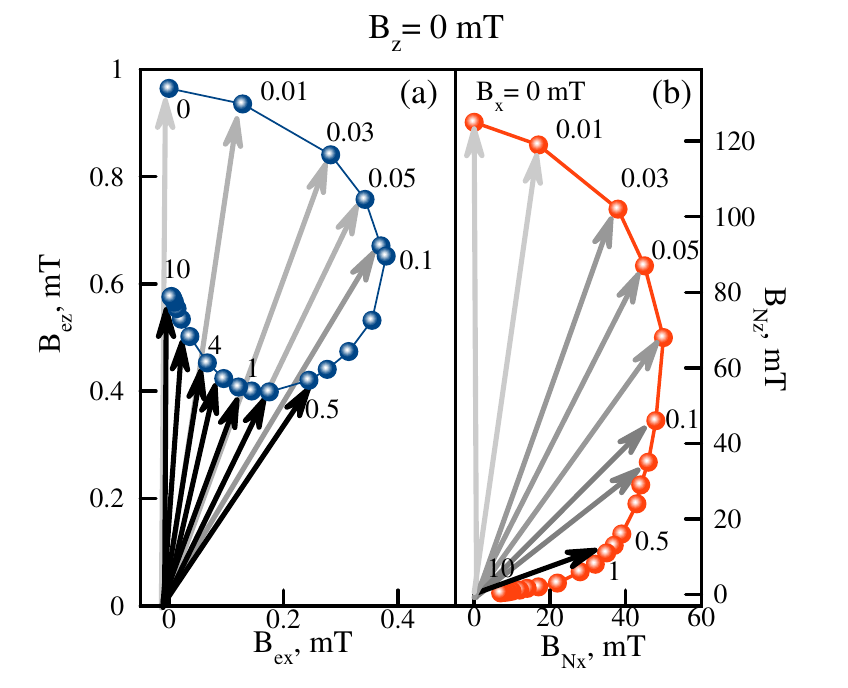}
\caption{Evolution of Knight field $B_e$ (panel a) and of nuclear field $B_N$ (panel b) at $B_z = 0$ as function of external magnetic field, $B_x$. Values of $B_x$ (in mT) are given for some curves. The step between points is not constant. Arrows show respective $\mathbf{B_e}$ and $\mathbf{B_N}$ vectors.}
 \label{fig:Five}
\end{figure}

The good overall correspondence of the simulated and measured Hanle curves confirms the validity of the model developed. In the framework of this model, we can get a clear idea about the vector representation of the time-averaged electron spin and nuclear polarization in the system under study. Figure~\ref{fig:Five} schematically shows the evolution of the respective vectors under variation of the transverse magnetic field, $B_x$, and for zero  longitudinal field. For uniformity, the electron spin and the nuclear polarization are presented as effective fields, $B_e$ and $B_N$, respectively. The diagrams are shown only for positive values of $B_x$. For negative $B_x$, the $x$-components of vectors $B_e$ and $B_N$ are negative so that the diagrams are symmetrical relative to the vertical axes.

The nuclear field at zero $B_x$ is only controlled by the Knight field, which is directed along to $z$-axis.
When a small transverse magnetic field, $B_x\ll B_e$, is applied, the nuclear field deviates from the $z$-axis
so that its $x$-component becomes orders of magnitude larger than the magnetic field $B_x$. For example, $B_{Nx}\approx 50$~mT at $B_x = 0.1$~mT.
This is a clear illustration of the  ,,amplification'' of the external magnetic field by the nuclear field.\cite{OO} The electron spin polarization follows
the nuclear field, which becomes quickly tilted with magnetic field and depolarizes the electron spin. This behavior of the electron spin explains
the small width of the central peak of the Hanle curve. For a further increase of the magnetic field, the magnitude of the nuclear field rapidly drops so that $|\mathbf{B_N}| \leq |\mathbf{B_x}|$ at $B_x\geq10$~mT.

\begin{figure}
\includegraphics[clip,width=.85\columnwidth]{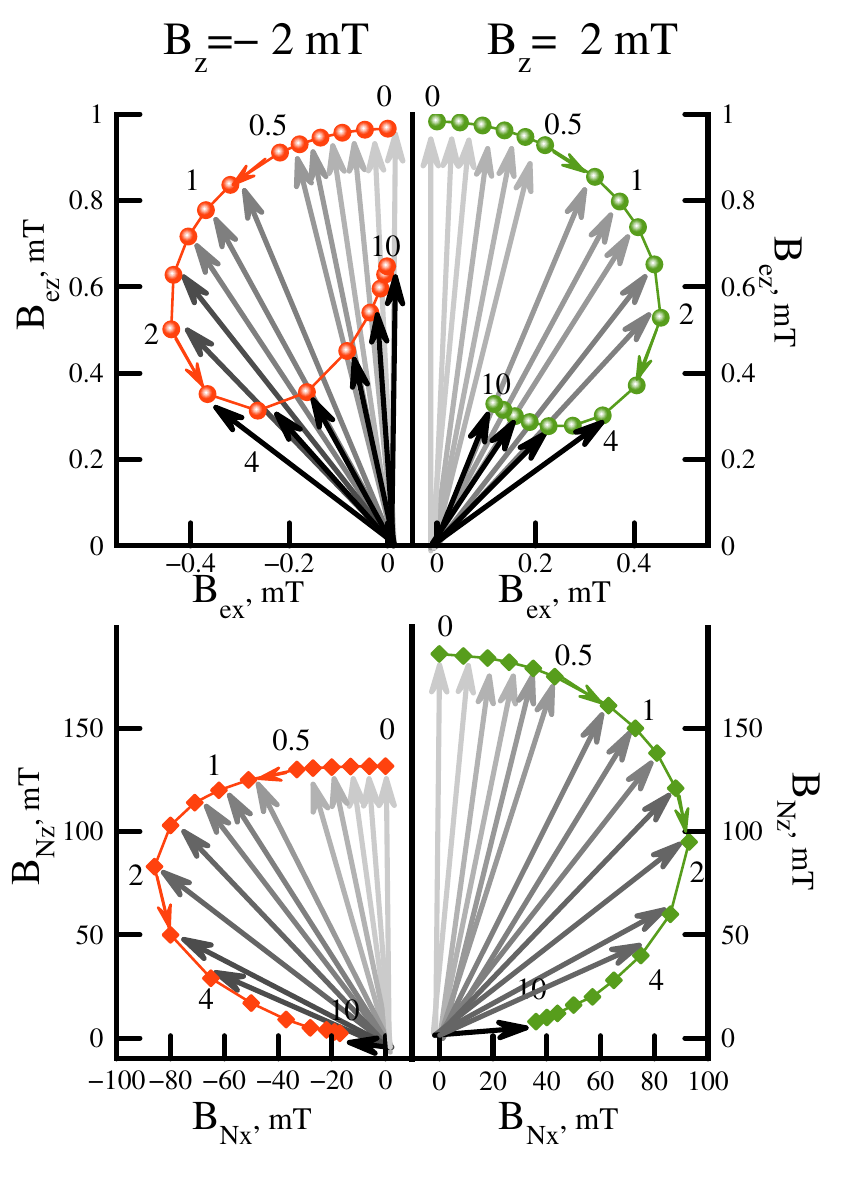}
\caption{Dynamics of the electron (upper panels) and nuclear (low panels) fields for application of negative (left panels) and positive (right panels) longitudinal magnetic fields with relatively large magnitudes.}
 \label{fig:Six}
\end{figure}

Application of a longitudinal magnetic field with magnitude larger than that of the Knight field significantly changes the behavior of the electron and nuclear polarization as demonstrated in Fig.~\ref{fig:Six}. An increase of the transverse magnetic field $B_x$ is accompanied by inclination and reduction of the Knight field; however the reduction is not as fast as at $B_z = 0$. The direction of the Knight field inclination depends on the sign of the longitudinal magnetic field, see left and right top panels in Fig.~\ref{fig:Six}. The nuclear field $\mathbf{B_N}$ is directed along the $z$-axis at zero transverse magnetic field and has the maximal value $B_N = b_N*S_0 = 200$~mT at positive $B_z$ when the Knight field and the longitudinal magnetic field add up (bottom right panel in Fig.\ref{fig:Six}. At opposite (negative) sign of $B_z$ when the fields are subtracted from each other, the total effective field acting on the nuclei is smaller, which results in some reduction of the nuclear polarization (bottom left panel in Fig.~\ref{fig:Six}). The direction of inclination of the nuclear field is also dependent on the sign of $B_z$. In particular, the $x$-component of the nuclear field is negative at negative $B_z$, so that compensation of the transverse magnetic field occurs at $B_x\approx10$~mT. This compensation results in partial restoration of the electron spin polarization and its reorientation again along the $z$-axis (see top left panel in Fig.~\ref{fig:Six}). The decrease of the magnitude of the Knight field relative of its initial value at $B_x = 0$~mT is the effect of the nuclear spin fluctuations, as discussed above.

\section*{Conclusion}

The experimental study of electron spin polarization in (In,Ga)As QDs as function of an applied transverse magnetic field for simultaneously applied small longitudinal magnetic field highlights a number of specific features of the hyperfine interaction in these systems. The analysis of experimental data has confirmed the prediction of Ref.~\onlinecite{MerkulovPRB02} about the significant influence of nuclear spin fluctuations (NSF) on the dynamics of the electron spin due to strong localization of the electron in QDs. The observed behavior is considerably different from that in extended semiconductor alloys studied in many works,\cite{OO} in which the electron density is spread out over a huge number of nuclei and the effect of the NSF, as a rule, is negligibly small.

We have restricted our analysis to the range of transverse magnetic fields  $|B_x| < 20$~mT, (see Figs.~\ref{fig:Two}) for two reasons: (i) the effect
of the  NSF is most prominent in this range, and (ii) the effect of the quadrupole splitting of the nuclear spin states is small.
However, the results of our calculations show that this effect cannot be totally ignored. In particular, it may be responsible for an anisotropy of the nuclear spin fluctuations and a rise of the wings of the Hanle curve. Another possible manifestation of the quadrupolar effect is the relatively large amplitude of the central peak for mutual compensation of the Knight field and the $z$-component of external magnetic field.

The consideration of the NSF-field has allowed us to quantitatively describe the Hanle curves in this range and to evaluate random and regular nuclear fields acting on the electron. We have also estimated the average magnitude of the Knight field and found that the magnitude of the nuclear polarization obtained from the simplified analysis based on neglecting the transverse component of the Knight field gives rise almost to the same result as the exact calculation. Therefore the simplified model can be useful for an express-analysis of the experimentally measured Hanle curves.

\section*{Acknowledgments}
This work was supported by the Deutsche Forschungsgemeinschaft, the Russian Foundation for Basic Research, and EU FET SPANGL4Q. The financial support from the Russian Ministry
of Education and Science (contract No. 11.G34.31.0067 with SPbSU and leading scientist A. V. Kavokin) is acknowledged.

\end{document}